\documentclass[]{aa}
\usepackage{rotating}   
\usepackage{graphicx}
\usepackage{txfonts}
\usepackage{natbib}  
\bibpunct{(}{)}{;}{a}{}{,} 

\newcommand{\percc}{cm$^{-3} $}        
\newcommand{\persc}{cm$^{-2} $}        
\newcommand{\kks}{K km s$^{-1} $}      
\newcommand{\kms}{km s$^{-1} $}        
\newcommand{\HH}{H$_2 $}               
\newcommand{\AMM}{NH$_{3} $}           
\newcommand{\AMMD}{NH$_2$D}            
\newcommand{\NTHP}{N$_2$H$^+ $}        
\newcommand{\NTDP}{N$_2$D$^+ $}        

\newcommand{\micron}{$\mu$m)}

\begin{document}

\title{Observing the gas temperature drop in the high-density nucleus of L~1544.\thanks{Based on observations carried out with 
the IRAM Plateau de Bure Interferometer and the Very Large Array. 
IRAM is supported by INSU/CNRS (France), MPG (Germany) and IGN (Spain). 
The National Radio Astronomy Observatory is a facility of the National Science Foundation operated under cooperative agreement by Associated Universities, Inc.}}

\author{Antonio Crapsi\inst{1,2}  \and
	Paola Caselli\inst{1,3}  \and
	Malcolm C. Walmsley\inst{1}  \and
	Mario Tafalla\inst{4}  }
   \offprints{Antonio Crapsi, \email{crapsi@arcetri.astro.it}}

\institute{
 Osservatorio Astrofisico di Arcetri, Largo E. Fermi, 5, I-50125, Firenze, Italy 
   \and
 Leiden Observatory, Leiden University, P.O. Box 9513, NL-2300 RA, Leiden, The Netherlands
 \and
 Harvard-Smithsonian Center for Astrophysics, 60 Garden St., MS 42, Cambridge, MA 02138, USA
  \and
 Observatorio Astron\'omico Nacional (IGN), Alfonso XII, 3, E-28014, Madrid, Spain}

   \date{Received 05 April 2007 / Accepted  27 April 2007}

\titlerunning{L~1544 \AMM - \AMMD \ observations at high-resolution}
\authorrunning{Crapsi et al.}

\abstract
{The thermal structure of a starless core is  crucial 
 for our understanding of the physics in these objects
and hence for our understanding of star formation.
Theory predicts a gas temperature drop 
in the inner $\sim$5000~AU of these objects,
but  there has been no observational proof of this. 
}
{We performed VLA observations of the \AMM \  (1,1) and (2,2)
 transitions towards the pre-stellar core L~1544 in order to 
measure the temperature gradient between the high density core
nucleus and the surrounding core envelope. Our VLA observation for the first time 
provide measurements of gas temperature in a core with a resolution smaller than 1000~AU.
We have also obtained high resolution Plateau de Bure 
 observations of the  110 GHz $1_{11}-1_{01}$ para--\AMMD \ line in order to further constrain
the physical parameters of the high density nucleus.   
}
{We combine our  interferometric \AMM \ and \AMMD \ observations
with available single dish measurements  in order to estimate
the effects of  flux loss from extended components upon our data. 
We  have estimated the temperature
gradient using a model of the source to fit our data in the $u,v$ plane.
As the \AMM (1,1) line is extremely optically thick, this also
involved fitting a gradient in the \AMM \ abundance. 
In this way, we also measure the [\AMMD]/[\AMM] abundance ratio in
the inner nucleus. 
}
{We find that indeed the temperature decreases toward the core nucleus
from 12~K down to 5.5~K resulting in an increase of a factor of 50\% in the estimated density of the core from the
dust continuum if compared with the estimates done with constant temperature of 8.75~K. 
Current models of the thermal equilibrium can describe consistently the observed temperature and
density in this object, simultaneously fitting our temperature profile and the continuum emission.  
We also found a remarkably high abundance of deuterated ammonia with respect to the ammonia abundance 
(50\%$\pm$20\%), which proves the persistence of 
nitrogen bearing molecules at very high densities ($2\cdot10^6$~\percc ) and shows that high-resolution 
observations yield higher deuteration values than single-dish observations. The \AMMD\ observed transition, free of the optical depth
problems that affect the \AMM\ lines in the core center, are a much better probe of the high-density nucleus and, in fact, its map peak at the
dust continuum peak. Our analysis of the \AMM\ and \AMMD\ kinematic fields shows a decrease of specific angular momentum from the
large scales to the small scales. 
}
{}
\maketitle
\section{Introduction}
Dense starless cores are thought to be the forerunners of solar mass
 protostars and hence determining their physical and chemical evolution is
 an essential step  in our understanding of star formation. Thus, measuring
their physical characteristics (temperature,density, velocity field) should 
give
us insight into the initial conditions just prior to the collapse of a
protostar. 
Most interesting in this regard are the cores of highest central density
and column density which appear closest to the ``pivotal structure'' in
which the core is on the verge of instability.

 Recent observations of both the dust continuum and molecular lines have
allowed a selection to be made of the cores of highest central density
 \citep{wardthompson1999,tafalla2002,crapsi2005a}.
It turns out that such cores have a few other properties in common such
as high deuterium fractionation and (in a few cases) some evidence for
increased velocity dispersion indicative of contraction in the
high density core nucleus.  Deuterium fractionation potentially gives
us important information on the ionization fraction   at high densities
while the observed velocities and line widths yield precious information
on angular momentum and infall. To exploit these to the full however
requires high angular resolution measurements in a nearby core. 
With this in mind, we decided to obtain high quality interferometer maps
of the nearby Taurus core \object{L~1544}
in the 23 GHz (J,K)=(1,1) and (2,2) lines of ammonia (\AMM)as well as in the 
 110 GHz $1_{11}-1_{01}$ transition of deuterated ammonia  (para--\AMMD).

The choice of ammonia and \AMMD \ for this study was partly based on the
fact that it has been found \citep[e.g.][]{tafalla2004} that nitrogen containing species 
such as ammonia and its deuterium isotopologues trace preferentially the
densest gas in nearby cores. The reason for this is not precisely known
but is likely to be related to the relative volatility of both atomic
and molecular nitrogen.  It is also important that the relative population
of the (1,1) and (2,2) levels of ammonia is an excellent temperature indicator
\citep[see e.g.][]{ho1983,danby1988,ungerechts1986}  due to the fact
that radiative transitions between the K=2 and K=1 ladders of \AMM \
are forbidden. We can therefore use the relative optical depths in
these two lines as a measure of temperature. 
One of the basic questions
about pre-stellar cores is whether they have  a positive temperature 
gradient outwards \citep[see e.g.][]{leung1975,evans2001,zucconi2001,galli2002,lesaffre2005}
with temperatures of the order of 6~K in their center, as one would expect
from theoretical considerations. 
On large scales $>0.05$~pc, there is already evidence
for higher {\rm dust} temperatures  on core edges than in core centers  
\citep{wardthompson2002,pagani2004,kirk2005} but
on the scale of 0.01~pc (15\arcsec \ at 140~pc), the situation
is less clear.  With this in mind, we obtained
and present in this work D-array VLA \AMM \
observations with a roughly 4\arcsec synthesised beam .
  We are thus sensitive to compact structure 
of size roughly 1000 AU. 

 Measuring the temperature gradient is of importance mainly as a
test of our understanding of the thermodynamics of pre-stellar
cores though it also is of some importance for the core stability
 \citep[see][]{galli2002}.  The temperature measurements test our
understanding of dust--gas coupling and CO cooling 
\citep[see e.g.][]{bergin2006}.  They are also important 
  for the interpretation of dust continuum emission maps 
where it is usual  to assume 
an isothermal dust  temperature and spherical symmetry in order
to derive the density distribution.  The latter is often not valid
as we discuss later  \citep{goncalves2004,doty2005} but is
 nevertheless a useful zero order approximation.  Detailed
modelling of L~1544 has been carried out by \citet{doty2005} 
who fit the mm--submm continuum  data 
of that source with a dust temperature drop
from  15~K to 5.4~K.  We note however that the ammonia 
single dish data for several sources 
including L~1544 show no evidence for
gradients  \citep{tafalla2002,tafalla2004}  and thus the
existence of temperature gradients in  pre-stellar cores
is uncertain.

L~1544 is a well-studied low-mass starless core in Taurus, which presents
a large central density ($> 10^6$ \percc),
 considerable CO freeze-out ($>$ 93\%) 
and high degree of deuteration (N$_2$D$^+$/N$_2$H$^+$ $\sim$0.25).  
These properties, combined with kinematical
probes of infall (differences in velocity between thin and thick tracers,
and increasing line widths of thin tracers toward the core centre) makes
this core a good candidate for being
on the point of becoming unstable \citep{crapsi2005a}.  One expects
the dynamics of the inner core to be better sampled by deuterated
species and we have therefore 
obtained high resolution observations of the $1_{11}-1_{01}$
 transition of para--\AMMD  at 110 GHz with
the IRAM Plateau de Bure interferometer (PdBI).
The resolution is rather similar to that of
the VLA  and we are thus able to measure the
deuterium fractionation in the core nucleus. We also can derive
important limits on angular momentum and infall. 

This paper is organized as follows, in section~\ref{obs} we report the details of 
the VLA and PdBI observing runs, in section~\ref{res} we present our basic results, 
in section~\ref{ana} we describe our analysis procedure and in
section~\ref{concl} we summarize our results.

\section{Observations}\label{obs}
 We used the Very Large Array (VLA) to observe L~1544 in the (1,1) and (2,2) transitions of para-\AMM \  
 (at 23694.4949~MHz and 23722.6349~MHz respectively) and the Plateau de Bure interferometer (PdBI) to observe the 
 ($1_{1,1}-1_{0,1}$) transition of para--\AMMD \  (at 110153.587~MHz). 
 The reference position we used in this work is the dust continuum peak taken from the 1.2~mm map of 
 \citet{wardthompson1999}:  (05:04:17.230,   25:10:47.70) J2000.0.

\subsection{VLA Observations}
 Observations at the VLA were performed in compact configuration (D) on March 8$^{th}$, 2003 and April 3$^{rd}$, 2003 
 for a total of 20 hours of observations.
 On-source observations were interleaved by the quasar 
 0510+180 (1~Jy at 1.3~cm according to the VLA calibrator  manual\footnote{http://www.aoc.nrao.edu/$\sim$gtaylor/csource.html}) 
 for phase calibration, whereas absolute flux calibration was performed using 3C286 (2.59~Jy at 1.3 cm). 
 The difference in average flux density  for 0510+180 between the two observing runs was $\sim$15\% and 
 we take this as the uncertainty in flux calibration.
 The \AMM (1,1) and (2,2) data were acquired simultaneously using the 2 IF spectral line mode, sampling 64 channels over
 two 781~kHz (or 9.8~\kms) bandwidths and reaching a channel width of 12.2~kHz (or 0.154~\kms).
 While the (2,2) band was centred at the main hyperfine rest frequency, the (1,1) band was offset by 306~kHz to include 
 both the main group of hyperfines and one secondary group
(F$_1$=2-1 at 23693.905~MHz)  of hyperfines.
 The VLA primary beam at the ammonia frequencies is 2\arcmin , while
 the synthesized beam, after applying natural weighting, reached 4\farcs34$\times$3\farcs45 with a position angle (PA)
  of 79\fdg9.
 In order to increase the signal-to-noise ratio we reconstructed the clean map of the \AMM(2,2) using a circular clean beam
 with full width at half power of 8\arcsec .
 Calibration and data reduction were performed using the Astronomical Image Processing System (AIPS) of NRAO.

\subsection{IRAM Plateau de Bure Observations}
 Plateau de Bure observations were performed in the C and D configurations (the most compact ones) on September 29$^{th}$ 2002 
 and December 18$^{th}$ 2002 for a total of 16 hours of
 observing time. The correlator setup was chosen to simultaneously observe the dust continuum at 110~GHZ and 230~GHz
 with a 1~GHz bandwidth and the para--\AMMD(1,1) and CO(2-1) with a resolution of 0.106~\kms \ and 0.051~\kms \ respectively. 
 Phase calibration was performed using 0528+134 and 0415+379 while absolute calibration was obtained using 3C454.3 for 
 the D-configuration run and 3C273 for the C-configuration.
 The two phase calibrators have a 20\% difference in flux in the two observing runs and we take this as the calibration
 uncertainty.
 Cleaning was performed with uniform weight and the resulting synthesized beam is 5\farcs8$\times$4\farcs5 in size and
 81\fdg4 in PA at 3~mm and 2\farcs5 by 2\farcs2 with PA  81\fdg0 at 1mm. 
 The PdBI data were reduced and calibrated using the Continuum and Line Interferometer Calibration (CLIC) developed by
 IRAM.

\section{Results}\label{res}
 \subsection{Maps and spectra}
 \subsubsection{VLA Ammonia}
\begin{figure}[htbp]
\resizebox{\hsize}{!}{\includegraphics{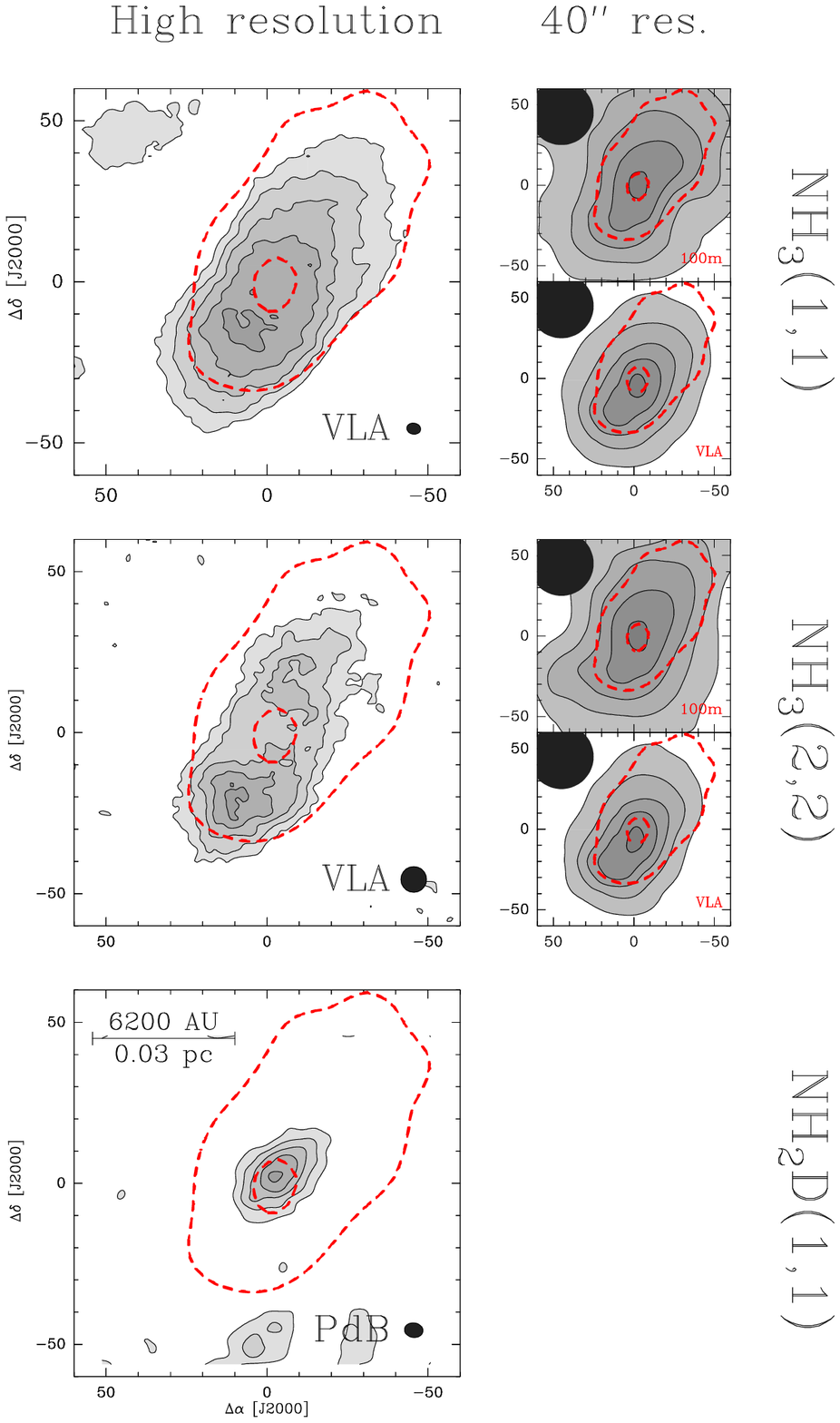}}
 \caption{ Integrated intensity maps of L~1544 in \AMM (1,1) and (2,2)\ and para--\AMMD($1_{11}-1_{01}$). 
 Dashed contours in all the panels are the 50\% and 95\% levels of the 1.2mm continuum taken from \citet{wardthompson1999} 
 and smoothed to 22\arcsec .
 {\bf Left column:} 
 Observations taken with VLA or PdBI were integrated over all the hyperfines available (\AMM(1,1) values 
 refer only to the main component and the first red component). 
 Angular resolution is 4\farcs34$\times$3\farcs45, 8\arcsec and  5\farcs8$\times$4\farcs5  for \AMM(1,1),  
 \AMM(2,2) and \AMMD(1-1), respectively,  and it is reported in the bottom right. 
 Levels are 15\% to 95\% by 20\% of the map peak which is 0.25, 0.05 and 7.0 \kks \ 
 for \AMM(1,1),  \AMM(2,2) and \AMMD(1-1), respectively. 
 {\bf Right column:} 
 Integrated intensity maps of the observations from the single dish (top) and interferometer 
 data smoothed to 40\arcsec (bottom). Levels are spaced percentually as in the left column but peak values are 8.5~\kks and 
 2.8~\kks for the single dish map of \AMM(1,1), and the smoothed VLA map of \AMM(1,1), and 
 0.47~\kks and 0.08~\kks for the \AMM(2,2) maps from Effelsberg and VLA. A 40\arcsec \ beam is reported at the top-left corner.}
 \label{Fmap}
\end{figure}

  In figure~\ref{Fmap} we show the integrated intensity maps of \AMM (1,1) and (2,2) from
  the VLA and the 100m antenna in Effelsberg \citep[from][FWHM $\simeq$ 37\arcsec]{tafalla2002}.
The qualitative difference between the two is striking: while the intensity measured with the single dish follows very closely 
the column density traced by the dust emission, the VLA maps of \AMM (1,1) peaks 20\arcsec south-east relative to the 
dust peak and the \AMM (2,2) emission shows a second peak also in the north west evocative of a ``hole'' in the emission. 
The angular size  of the ammonia integrated emission seen with the VLA is 75\arcsec$\times$36\arcsec \ in the (1,1) line
and 68\arcsec$\times$25\arcsec \ in the (2,2) line as compared to 105\arcsec$\times$58\arcsec \ in the dust emission.
 
 In order to properly compare  the interferometer and
single dish maps, we cleaned the VLA data with a 40\arcsec beam to simulate the 
  resolution of the 100-m data. The resulting map
 (Fig.~\ref{Fmap}) shows that 
  the two data sets are compatible.
 The discrepancy arises because the interferometer is insensitive to 
  large scale emission and traces preferentially
 compact high density gas clumps. 
  The peak intensities of  these maps are a measure of
 the amount of flux lost in the VLA observations
  due to extended structure: 68\% in the \AMM (1,1)
(70\% and 64\% for the main hyperfine component and the first red hyperfine component, respectively)
 and more than 80\% in the (2,2).
 As shown below the difference between these two is
significant and  thus the  (2,2) emission appears to be more
resolved out than the (1,1). This by itself is an indication that
the  temperature of the more extended (presumably lower density) 
material in the L~1544 core is higher than in the compact nucleus.

 However, an important characteristic of our results is that the
peak of the VLA (1,1) emission (effectively the ground state of
para \AMM) is clearly not coincident with the peak of the dust
emission. We believe this is mainly due to the fact that the
(1,1) emission is highly optically thick and thus our measurement
towards the peak of the main (1,1) satellite reflects the 
temperature at the surface corresponding to unity optical depth.

\begin{figure}[htbp]
 \resizebox{\hsize}{!}{\includegraphics{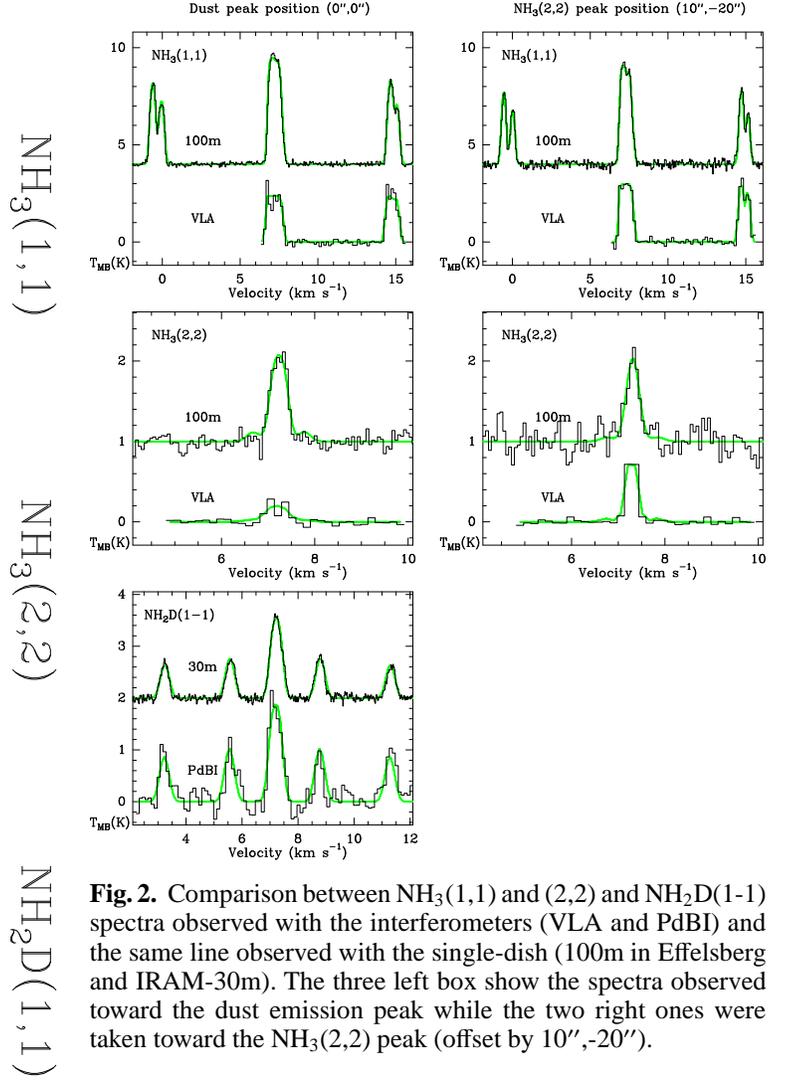}}
 \caption{ Comparison between \AMM (1,1) and (2,2)  and \AMMD(1-1) spectra observed with the interferometers (VLA and PdBI)
 and the same line observed with the single-dish (100m in Effelsberg and IRAM-30m).
 The three left box show the spectra observed toward the dust emission peak while the two right ones were taken 
 toward the \AMM (2,2) peak (offset by 10\arcsec,-20\arcsec).} 
 \label{Fspe}
\end{figure}

This is substantiated by Fig.~\ref{Fspe} where we compare
the spectra from the  VLA  with those observed with the single dishes.
We measure at the dust peak a ratio of the satellite F=2-1 emission to 
the main (F=1-1 and 2-2 blend) of 0.72 as compared to the expected 
ratio of 0.278 in the optically thin limit. This  \citep[see e.g.][]{ho1983}
allows an estimate of the main line optical depth 
and we derive a higher value from the VLA data (with an optical depth integrated over all the hyperfine components of 
$\tau_{tot} > 30$) than from the 100m 
($\tau_{tot} = 8.6$), while 20\arcsec \ further from the peak
the difference in opacity 
is reduced ($\tau_{tot} = 20$ vs. $\tau_{tot} = 7.7$) and
a smaller amount of flux is resolved out (we recover  75\% of the  flux). 
Our optical depths are in fact so high that they may affect the temperature gradients
discussed later.
These issues will be treated more quantitatively in section~\ref{tempMC} where we 
report our radiative transfer calculations combined
with the $u,v$ modelling of the emission. For the moment, we note however
that these results are consistent with a high ammonia column density in
the neighbourhood of the dust peak. 
  
\subsubsection{\AMMD \ with Plateau de Bure}
Our \AMMD \ data are extremely useful in that they confirm that
ammonia  remains in the gas phase at the densities of order $10^6$
\percc \ close to the dust peak. As in the case of ammonia, one can
use the relative intensities of the hyperfine satellites to provide
a measure of optical depth and in this case (see fig.~\ref{Fspe}) we
find intensity ratios of the main component versus the first red component (=0.3) consistent  with 
optically thin LTE (=0.278).  Thus the fact that the \AMMD \ intensity 
(see Fig.~\ref{Fmap}) peaks at the dust peak can be considered
a demonstration of the fact that \AMMD \ is a tracer of the dense
gas in the vicinity of the dust peak.  It can thus be used  to
measure the  kinematics in this region.  The angular size of
the \AMMD \ emission seen by the interferometer is 27\arcsec$\times$12\arcsec \ as compared
to the dust emission size of 105\arcsec$\times$58\arcsec . 
Comparing the intensity at peak observed by the PdB with an IRAM-30m spectrum taken towards the nucleus of L~1544
we measure that 50\% of the \AMMD \ total flux is recovered by the interferometer.

 \subsection{Kinematics}
We   used the interferometric data in order to
fit the    \AMM(1,1)  and (2,2) and the \AMMD  \ spectra . 
In this manner, we obtained maps of velocity and line-width of the various
tracers.  Our results here are sensitive  to
differential flux loss in different channels  as well as to the
effects of opacity but, for the reasons discussed above, these problems
are  unlikely to be of great importance for \AMMD . 
The central velocity information was used to compute the velocity
gradient across the core, in analogous fashion to
\citet{caselli2002a}  using \NTHP \ and \NTDP .

\begin{figure*}[htbp]
\resizebox{\hsize}{!}{\rotatebox{-90}{\includegraphics{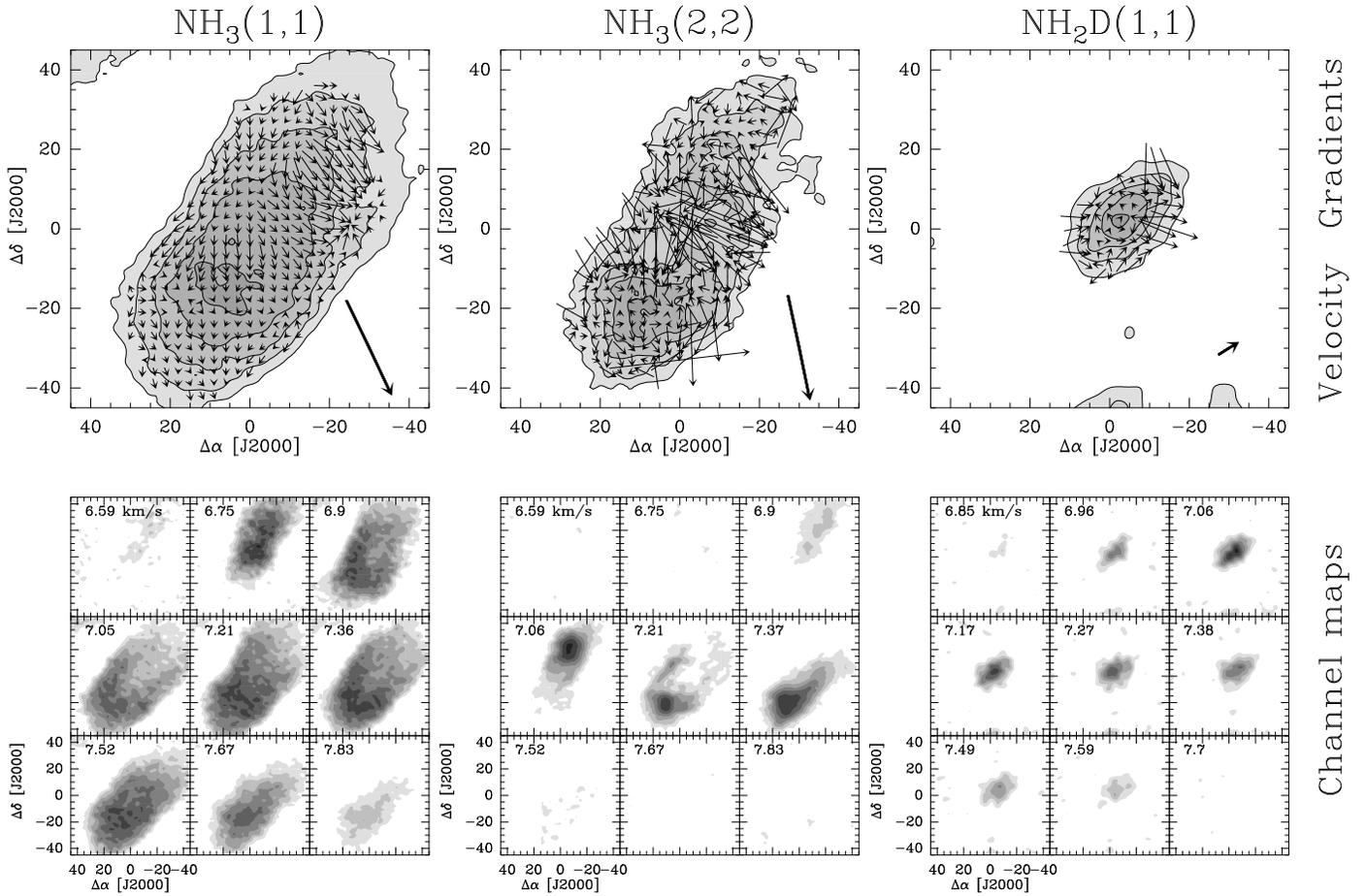}}}
 \caption{ Kinematics in L~1544.  {\bf Top row:} velocity gradients calculated in 9 adjacent points are overlaid on the integrated
 intensity map for each observed transition. 
 The direction of the arrow points to the increasing velocity while the lenght measure the magnitude of the 
 gradient (a 10\arcsec \ arrow corresponds to a gradient of 33~\kms~pc$^{-1}$).
 The thick arrows in the bottom-right represent the total gradient evaluated with all the fits (the magnitude for those is 
 ten times smaller: i.e. a 10\arcsec \ arrow corresponds to a gradient of 3.3~\kms~pc$^{-1}$) 
 {\bf Bottom row:} channel maps of the main component of each observed line (VLA and PdBI). }
 \label{Fcmap}
\end{figure*}

The velocity maps derived in this fashion are shown in
Fig.~\ref{Fcmap}.  \AMM(1,1)  and (2,2) show the same general
pattern with a total gradient of $\sim$9~\kms~pc$^{-1}$  
with a PA of 160 degrees west of north.
This pattern can be seen also in the channel maps 
presented in Fig~\ref{Fcmap}: in both lines
the emission is brighter in the north at blue velocities 
to become stronger in the south for higher velocities.
Over a scale of 1 arc minute(0.04 parsec), this amounts to a
specific angular momentum of 0.015 pc km s$^{-1}$ or
$5\, 10^{21}$ cm$^{2}$s$^{-1}$.  
Our results match the velocity gradients calculated by \citet{caselli2002a} from single--dish 
observations  \NTDP , showing gradients of $\sim$6~\kms~pc$^{-1}$ with a similar position 
angle as \AMM .

The different morphology of the \AMMD \ maps compared to those of \AMM \ already tells us that the two species
traces different material, this is confirmed in the velocity maps. 
in fact, the \AMMD \ velocity gradient map shows no clear sign of
rotation with  total gradient of $\sim$2~\kms~pc$^{-1}$  
with a PA of 60 degrees west of north.  The corresponding specific
angular momentum is 0.001 pc km s$^{-1}$ or $3.5\, 10^{20}$ cm$^{2}$s$^{-1}$, about an order of magnitude lower than the
specific angular momentum measured with \AMM . We note that the
position angle obtained from \AMMD \ is different from that seen in  \NTDP \ by \citet{caselli2002a} and may suggest a difference between 
the bulk motions of neutral and ion species in the inner core.

We note that the specific angular momentum obtained from \AMM \ is typical of starless cores \citep[see e.g.,][]{ohashi1999}, while the smaller
specific angular momentum derived from \AMMD \ is closer to the values of protostellar envelopes. Observations with higher velocity resolution should 
be needed to confirm this loss of specific angular momentum towards the small scales; however, this differential rotational properties between 
the inner envelopes and the ambient cloud is consistent with other observations in more evolved objects \citep{belloche2002}.

 \subsection{Dust Continuum and CO(2-1)}
  The Plateau de Bure Interferometer observations, together with the \AMMD(1,1) , simultaneously 
  mapped L~1544 in the dust continuum at 230~GHz and 110~GHz, and in CO(2--1).
  These observations resulted in $3\, \sigma$  non-detections of  0.9~mJy~beam$^{-1}$ and 0.5~mJy~beam$^{-1}$ 
  for the 1~mm  and 3~mm continuum  respectively. The corresponding limit for CO(2--1) was 80~mJy~beam$^{-1}$ 
  for the CO(2--1).

  Dust and CO emission from  L~1544 is easily detected by 
single dish telescopes \citep{tafalla2002}, but their 
  distribution is clearly too extended even for the most
 compact PdBI configurations leading to a total 
  loss of flux. 
  The absence of {\it compact} continuum emission at 1~mm confirms the starless nature of L~1544
and for an (arbitrarily) assumed temperature of 30~K yields an upper limit to the mass of
a  protoplanetary disk (on size scales of 1 beam i.e. 2\farcs5 by 2\farcs2) of $1.8 \cdot 10^{-4}$ solar masses.
This estimate is a strong upper limit because we assume the small dust opacity  for 
grains coated with thin ice mantles and maximum size of 0.25~\micron \ \citep[equal to 0.78 cm$^{2}$~g$^{-1}$][]{ossenkopf1994}; 
if the maximum size of the dust grains in the disk grows to 1~mm the upper limit should be decreased by a factor of four 
\citep[e.g. see dust opacities from][]{draine2006}. 
  In fact, if L~1544 did contain a low-luminosity protostellar object
 similar to  that found by the  Spitzer Space Telescope
  in \object{L~1014-IRS} \citep{young2004}, then we might have  observed the continuum from the disk
surrounding the protostar (expected to be of few mJy in L~1014-IRS at 1~mm, \citealt{young2004}).
 We assume here low optical depth conditions appropriate for large ($>100$~AU) disks observed at millimeter wavelengths.
For very small optically thick disks, our result amounts to a limit  of  $\rm T_{disk} < 50 \, K \, (R_{disk}/1 AU)^{-2}$. 
 In any case we would have certainly observed a molecular outflow similar to that in 
 L~1014-IRS  (which was detected by \citealt{bourke2005} with a R.M.S. 
of 80~mJy~(1\arcsec beam)$^{-1}$,   i.e. 4.8 times less sensitive than the present PdBI observations). 
\\
Our data thus confirm the {\it starless} nature of L~1544.

\section{Analysis}\label{ana}

\subsection{Evaluation of temperature in LTE}\label{tempLTE}
 Deriving temperatures from \AMM \ data has mostly been done
using the so--called ``standard analysis''  \citep[e.g.][]{ungerechts1986,stutzki1985}. 
One can use for example equation A7 from \citet{ungerechts1986} to
compute the ``rotation temperature'' $T_{12}$ defined by the ratio of
populations in then (2,2) and (1,1) levels. It is well known
\citep{walmsley1983} that at low temperatures this is in
practice close to the kinetic temperature (i.e. to LTE between the
two levels). Thus,

\begin{equation}
T_{12} \, = \, T_{2,2}/\ln[\tau_{11}^{tot}/(0.45\, \tau_{22}^{tot})]
\end{equation}
    
 where $T_{2,2}$ is the excitation energy of the (2,2)
relative to the (1,1 level ($41$~K), and $\tau_{11}^{tot}$ and
 $\tau_{22}^{tot}$ are the total optical depths of the (1,1) and (2,2)
 levels respectively (i.e the sum over all satellites).
We assume here equality of line-widths and ``amplitude factors'' 
\citep[see][]{ungerechts1986}) for the two lines.

 We thus see that our accuracy in determining the
temperature depends on our accuracy in determining the optical depths.
This is limited in the case of the (1,1) by the fact that using the
relative intensities of the satellites becomes inaccurate for large optical
depths (ratios close to unity). In the case of the (2,2) on the other
hand, we do not measure the (2,2) satellites and thus are forced to derive
the (2,2) optical depth from the ratio of the (2,2) and (1,1) main lines. 
Thus, we have
\begin{equation}
\tau_{22}^{tot} \, = \, -1/ f_{22} \,
 \ln\left\{1-(T_{M,2}/T_{M,1}) \, \left[1-\exp(\,-f_{11} \tau_{11}^{tot}) \right] \right\}
\end{equation}
 where $T_{\rm {M},2}$ and $T_{\rm{M},1}$ are the observed (2,2) and (1,1) brightness 
temperatures of the main hyperfine components and  $f_{22}$ and $f_{11}$ are
the relative intensities of the main hyperfine structure for the (2,2)  and (1,1)
(0.5 and 0.8). 
The error-bars were evaluated propagating the errors on $T_{\rm M,2}$, $T_{\rm M,1}$ and $\tau_{11}^{tot}$.
Note that doubling the intensity ratio between the (2,2) and (1,1) transitions would result in an increase of temperature of only 0.5~K,
while doubling the opacity of the (1,1) transition would decrease the temperature of 0.7~K. We  expect that
neither the errors on the opacity determinations nor the differential flux loss
 between the (1,1) and (2,2)  reach this. Our main errors in this  approach
are caused by the breakdown of the isothermal assumption and to differential
flux loss between (2,2) and (1,1) (twice as much lost in (2,2)). 

\begin{figure*}[htbp]
\centering
\resizebox{\hsize}{!}{\includegraphics{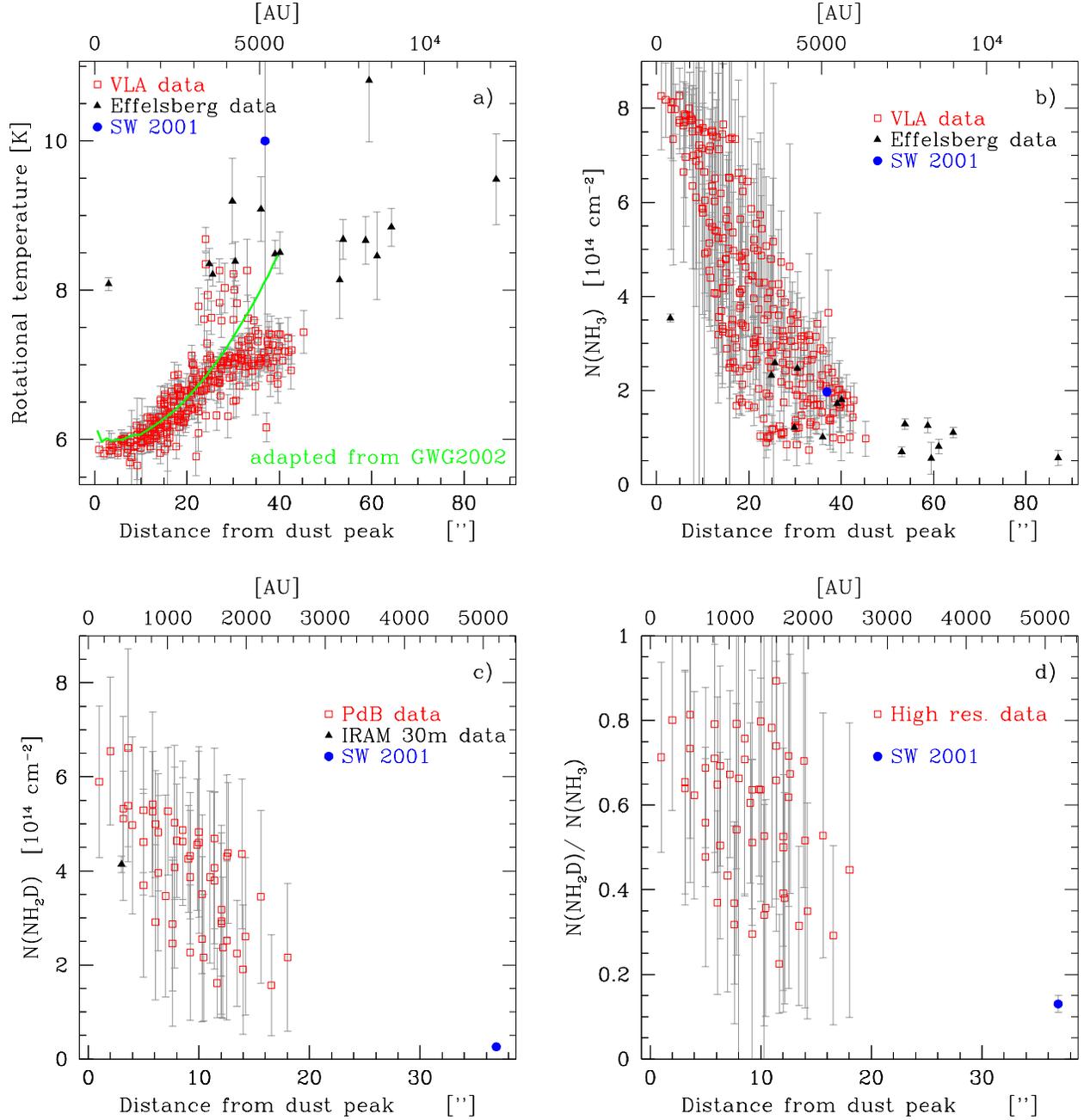}}\\
 \caption{ Temperature and column densities profiles as derived in the LTE approximation (see sections \ref{tempLTE} and \ref{coldensLTE}). 
 Observations from VLA are represented with empty squares, while those obtained at the 100m single-dish in Effelsberg are shown with filled triangles.
 Literature data from \citet{shah2001} are shown with filled circles.
  {\bf Panel a):} Gas temperature profile in L~1544 both from single-dish (triangles and circle) and VLA data (empty squares). 
  The solid line is an adaptation of the \citet{galli2002} model now using a central density of $2 \cdot 10^6$~\percc \ which is consistent 
  with the low temperature observed (see section \ref{tempLTE}).  
  {\bf Panel b):}  \AMM \ column density profile from both single dish data (triangles and circle) and VLA data (empty squares). 
  {\bf Panel c):}  same as panel b) but for \AMMD .
  {\bf Panel d):}  $N$[\AMMD]/$N$[\AMM] profile for the  literature data and the new high-resolution data.  }
 \label{Fpro}
\end{figure*}

 In panel a of Fig.~\ref{Fpro}, we present the temperatures derived using this
technique as a function of projected distance from the dust peak.
Both our present VLA data set and the 100-m results of \citet{tafalla2002}
are shown. 
For comparison, we show results from the theoretical model of
\citet{galli2002}.  
A clear drop in temperature is found toward the continuum peak 
on the basis of the VLA data in contrast to the single dish data
which show a rather flat profile. This is a clear suggestion 
that the compact high density core nucleus sampled with the interferometer
is at a lower temperature than its surroundings.

Our VLA data suggest that the gas temperature in the inner nucleus reaches values as low as 6~K.
 This contrasts with the minimum dust temperature of 11~K derived by 
\citet{wardthompson2002} from ISOPHOT observations between 90 and 200\micron , although it should be noted that these 
observations were done with $\sim$80\arcsec \ of angular resolution and that far infrared observations alone are more sensitive to
extended, warmer ($T>30$~K) dust coming from smaller optical depths, like the outskirts of starless cores.
In fact our observations are fully consistent with the best fit of \citet{doty2005} who modelled 
the dust continuum at 450, 850 and 1300~\micron \  (hence typical of cold dust) with a 3D
 radiative transfer code resulting in a central temperature of 5.4~K.

In Fig.~\ref{Fpro} we superimposed a modified version of the
 theoretical prediction (green curve) made by \citet{galli2002}. 
Modifying the model in \citet{galli2002} was necessary since this was
 calculated starting from a density profile that was derived from a 
constant dust-temperature of 8.75~K \citep{tafalla2002}.
Since the observed drop in gas temperature reflects a drop in
dust temperature (the two are well coupled at high densities 
and low temperatures, see e.g. \citealt{goncalves2004}), we expect the central density to be enhanced using this new information.
Since the temperature profile influences the estimate of the density profile and vice-versa, we had to 
iterate few times the following process; i) calculate the gas temperature from a given density profile ii) use that temperature profile 
to derive a new density profile fitting the 1.2~mm map iii) use this new density profile in point i).
We converged after few iterations with the model represented in panel a) of Fig.~\ref{Fpro}.
Note that this is {\bf not} a fit of the temperature data but just an iterative process to make the model of \citet{galli2002} self 
consistent with the 1.2~mm map.
Nevertheless the agreement with the data is very good in the limits of validity of the model and the data.
Including the temperature profile, the central density in L~1544 result to reach $\sim 2 \cdot 10^6$~\percc \ while the size
 of the flattened area shrinks by 30\% and the
slope of the density profile at large radii is not significantly changed.

\subsection{Column density and Deuteration in LTE}\label{coldensLTE}
 Large deuterium enrichment is a characteristic of the high density
nuclei of pre-stellar cores and it is a useful test of our understanding of 
the chemistry \citep[e.g.][]{aikawa2005,flower2006} to measure the
 [\AMMD]/[\AMM]  ratio at high angular resolution. We have therefore
combined our VLA and PdeB data for this purpose. 

We derived the column density of \AMM \ and \AMMD \ in the constant
 excitation temperature approximation \citep[CTEX, see also][]{caselli2002b}. 
In the case of \AMM , opacity and excitation temperature were derived 
through hyperfine structure fitting of the (1,1) transition using the
formalism discussed earlier. We assumed an ortho-to-para ratio (o/p) of 1 when doing this
and used a classical partition function at the temperatures of interest to correct for the excited states. 
The results from single-dish data can be compared to the
column densities evaluated by \citet{tafalla2002} using the more
sophisticated Monte Carlo approach. 
As shown in Fig~\ref{Fpro}, from the single dish data we derived a column density of \AMM \ of $3.5 \cdot 10^{14}$~\persc, this, 
combined to a column density of \HH \ of $9.4 \cdot 10^{22}$~\persc \ \citep{crapsi2005a}, furnishes 
an abundance of $3.7\cdot 10^{-9}$ comparable to \citet{tafalla2002}.
The interferometric observations give a factor of $\sim$2 larger
 column densities, which may  be an under--estimate in that 
our (1,1) optical depths are in some positions lower limits.

In the case of \AMMD , we are not observing the ground transition 
($1_{0,1}-0_{0,0}$ at 332.8~GHz) and thus we need an estimate  
of the ``rotation temperature''.  
We solved this problem evaluating  $T_{ex}$ from the ground state to
 the observed level ($1_{1,1}-1_{0,1}$) using
 Eq.~2.21 of \citet{winnewisser1978}:
\begin{eqnarray*}
\exp(-T_0/T_{ex})   & = & \exp(-T_0/T_{BB}) \times \\
  & & \times \frac{1+ (C_{ij}/A_{ij}) \exp(-T_0/T_k) [\exp(T_0/T_{BB})-1] }{1+(C_{ij}/A_{ij}) [1-\exp(T_0/T_{BB})]} 
\end{eqnarray*}
where $T_0$ is the energy between the levels i and j,
 $T_{BB}$ is the 2.7~K background, $T_k$ the kinetic temperature,
 $A_{ij}$ is the Einstein
coefficient and $C_{ij}$ is the collision rate between the levels $j$ and $i$.
The latter can be expressed as  $C_{ij}=\sigma_{ij} \left<v_{th}\right> n(\rm H_2) $, where $\sigma_{ij}$ is the cross section of the $i-j$
transition  ($=8 \cdot 10^{-16}$~\persc, cf. \citealt{green1976,turner1978}), $\left<v_{th}\right>$ is the thermal velocity dispersion and $n$(\HH) the \HH \ volume density.
Taking the kinetic temperature from the \AMM \ data and the $n$(\HH) from the 1.2~mm data we were able to evaluate the excitation temperature of the
ground transition and thus the partition function of \AMMD .
This amounts to   ignoring the splitting of the J=1 level and is
a crude procedure but we think it gives a reasonable estimate of the 
fraction of \AMMD \ in the ground state. We assumed an o/p ratio of 3 for \AMMD  , consistent with the average observed \AMMD \ 
o/p ratio in starless cores (\citealt{shah2001}, but see also \citealt{flower2006} for a theoretical approach).

The column density in the level ($1_{1,1}-1_{0,1}$) was instead calculated in the CTEX limit.
The profile  of $N$(\AMMD)  is shown  in panel  c) of Fig~\ref{Fpro}.
In this case the column density derived from the interferometry data 
is only slightly higher than the estimate from the single dish. 
Dividing N(\AMMD) by N(\AMM) we were able to evaluate the degree of
D-fractionation in \AMM at the dust peak of L~1544 
(see panel d) of Fig.~\ref{Fpro}).
The abundance ratio [\AMMD]/[\AMM] \ was found to be 0.5$\pm$0.2,
 thus  higher than in \NTHP \ \citep[0.26,][]{caselli2002b}, although caveats in the different flux loss between VLA and PdB apply.

\subsection{Model Fits of the Temperature and \AMM \ abundance }\label{tempMC}

In order to account for  the temperature gradient along the
line of sight as well as the complex flux loss from the interferometer, 
we  made a model of the temperature, density and
 abundance structure of the core. We then modelled
the excitation of the molecule and the radiative transfer of
 the emission, and finally
sampled the synthetic data cube in the $u,v$ plane to compare with
 the VLA observations.

The excitation and radiative transfer calculation were performed using
the 1D Monte Carlo radiative transfer code for molecular lines written by \citet{michiel2000}. 
The code takes as input for each radial cell: the density of \HH , the abundance of the molecule of interest, the dust and gas temperature and
the radial velocity. Then, using the collisional rates from \citet{danby1988} and the energy levels, transition frequencies, Einstein A coefficients 
(from JPL)\footnote{these informations are collected in the Leiden Atomic and Molecular Database at http://www.strw.leidenuniv.nl/$\sim$moldata/}
we can evaluate the population in each energy level and in each cell. This result
is taken as input for the ray-tracing module to produce a data cube of the synthetic emission.
We modified the ray-tracing module to include the hyperfine structure of \AMM .
The output data cube can then be sampled in the $u,v$ plane to reproduce the VLA observations. 
We produced such synthetic visibility datasets with the MIRIAD task UVMODEL.
The Monte Carlo code used in this paper was carefully tested against the one used by \citet{tafalla2002} resulting in a very good agreement. \\

We parametrized the structures of density, temperature and abundance in the following way:
$$n(r) =\frac{n_0}{1+( r/r_{0,n})^{\alpha_n}}$$
$$X_{NH_3}(r)=X_0 (n(r)/n_0)^{\alpha_X}$$
$$T(r) =T_{out}-\frac{T_{out}-T_{in}}{1+( r/r_{0,T})^{1.5}}$$

 The forms for the density and abundance profiles where chosen for
  consistency with the results of \citet{tafalla2002}. The form chosen
  for the temperature structure has mainly the virtue of simplicity.
  It is consistent with those computed by \citet{evans2001}
  (e.g. that shown in their Fig.6 can be represented by $T_{out}=13.1$~K, $T_{in}=6.5$~K and $r_{0,T}=26$\arcsec)
  though we note that such models neglect the breakdown of gas-grain coupling in the outer parts of the core.
We therefore ran the calculation for  a grid of values for
 $T_{out}$, $T_{in}$, $r_{0,T}$,  $X_0$, $\alpha_X$ and compared
 the results with  the observations of the \AMM (1,1) main and first hyperfine components and the \AMM (2,2) main component observed at VLA and Effelsberg.
The density distribution is not constrained by the \AMM \ observations but by the 1.3~mm dust continuum.
For each temperature distribution we then recomputed the density distribution that would fit the 1.3~mm observations in similar fashion to
\citet{tafalla2002}.

The quantities we wish to match with both the VLA and the Effelberg observations are 
\begin{itemize}
\item the integrated amplitude vs. $u,v$ distance profile for each of the components (\AMM (1,1) main and first hyperfine (hf) and \AMM (2,2) main) 
\item  the profile of the ratio between \AMM (2,2)$/$ \AMM (1,1)$_{hf}$ (which roughly speaking is a measure of the temperature)
and the ratio between \AMM (1,1)$_{main}/$ \AMM (1,1)$_{hf}$ (which roughly speaking is a measure of the column density)
\item the line profile of all the observed data
\end{itemize}
We compared the amplitude profile both versus the $u,v$ distance of the baselines and versus the distance to the centre of the source in the inverted 
image\footnote{for this comparison we did not "clean" either the data or the model since this would 
have introduced further uncertainties; comparing the "dirty" maps is instead exactly equivalent to the comparison in the $u,v$--amplitude plane.}.
Those plots are equivalent since they differ only for a Fourier transform, however since the plot of amplitude versus distance from the source centre shows better 
the emission in the centre of the source we choose to present only those plots.

\begin{figure}
\centering
\resizebox{\hsize}{!}{\includegraphics{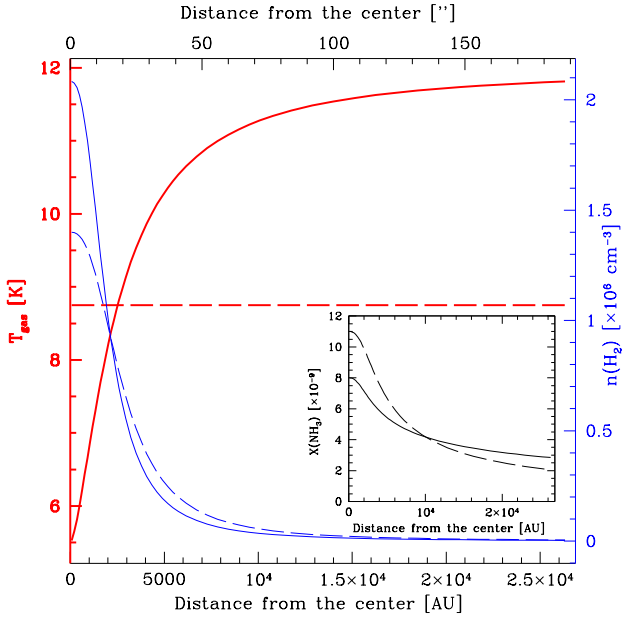}}\\
 \caption{Temperature, density and \AMM \ abundance profile  that best fit the interferometric and the single dish observations simultaneously (solid lines).
  The profiles relative to the best fit to the single dish data alone by \citet{tafalla2002} are reported in dashed lines for comparison.
   }
 \label{Fmod}
\end{figure}

\begin{figure*}[htbp]
\centering
\resizebox{\hsize}{!}{\includegraphics{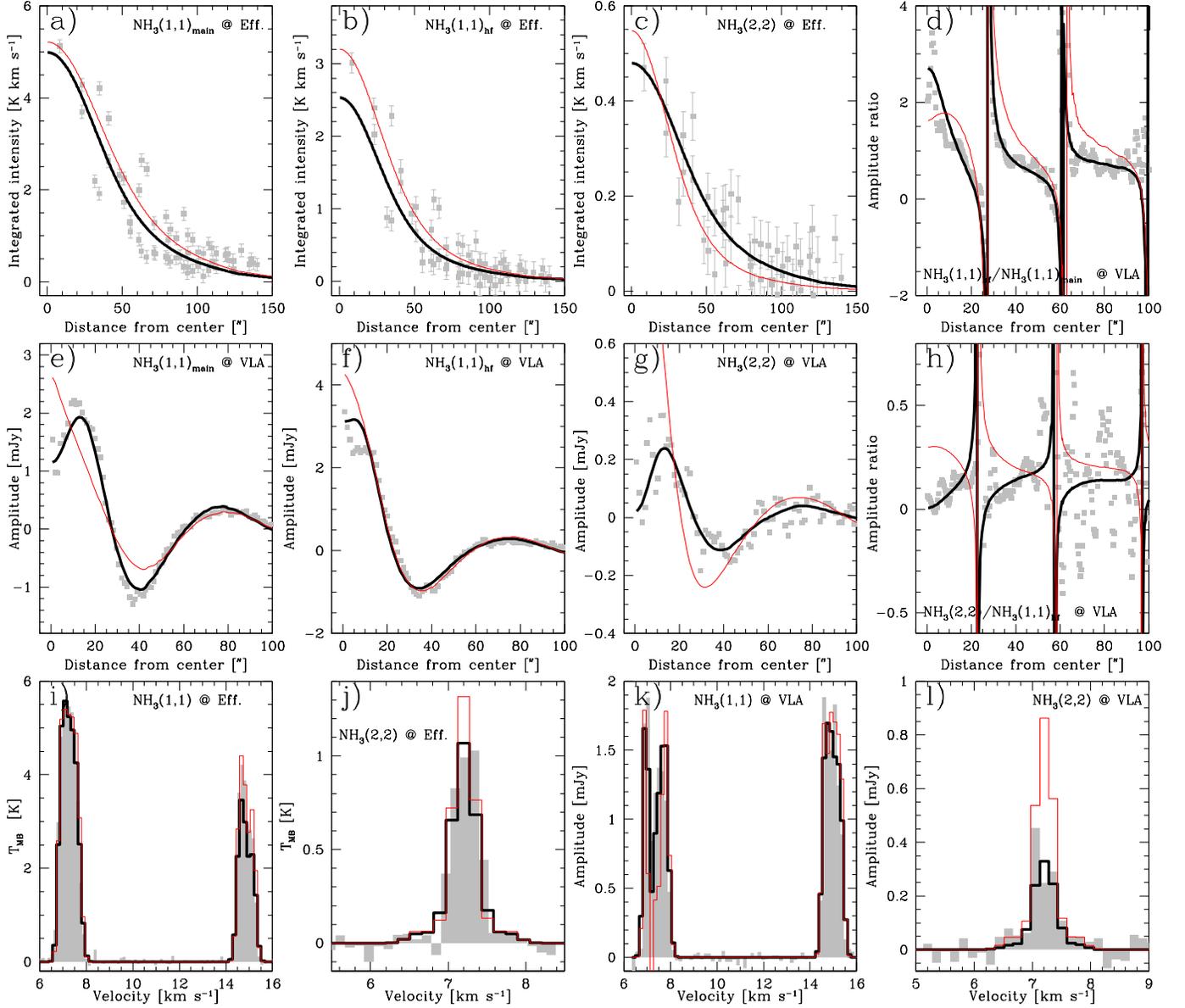}}\\
 \caption{Comparison between the observations of \AMM (1,1) and (2,2) taken with the 100m antenna of 
 Effelsberg and the VLA, and the model prediction.
 Data are always displayed in light grey. The best fit is shown in bold black lines, while the thin red curves show the best fit of the single dish data alone. 
 In panel a) we show the integrated intensity of the main hyperfine component of \AMM (1,1) as 
 observed with the single-dish. Panel b) and c) are similar to panel a)
 but for the first red hyperfine component of \AMM (1,1) and the main component of the \AMM
 (2,2). Panels d) to h) present the intensity profiles observed with the VLA compared with the 
 models. Both the data and the models were inverted from the $u,v$ plane to the image plane 
 using a fast fourier transform and then azimuthally averaged; the consequent negative values 
 are not due to errors but to the negative side-lobes, and we chose to include them in the
 modelling rather than add further uncertainties due to cleaning. 
 In panel d), we show the intensity ratio profile between the first red hyperfine component 
 of \AMM (1,1) and the main component, this is an indicator of column density distribution; the 
 asymptotic behaviour at 25\arcsec \ and 60\arcsec \ is due to the different extension of the two maps that determine a different size of the negative
 side-lobes. 
 In panel e), f) and g) we present the intensity profiles of the \AMM (1,1) main component, 
 \AMM (1,1) first red hyperfine and  \AMM (2,2) main component, respectively. Panel h) shows the amplitude 
 ratio profile between the \AMM (2,2) main component and the first red hyperfine component 
 of \AMM (1,1), this is an indicator of temperature.
 From panels i) to l) we compare the spectra at peak observed with the single dish and the interferometer with
 the models.}
 \label{Fmc}
\end{figure*}

The model that was found to best fit all this quantities has 
$T_{out}=12$~K, $T_{in}=5.5$~K, $r_{0,T}=18$\arcsec, 
$X_0=8 \cdot 10^{-9}$, $\alpha_X=0.16$. 
The density, temperature and abundance profiles of this model are shown in fig.~\ref{Fmod} in solid lines. As a comparison we show  with dashed lines 
the profiles relative to the best fit of the single--dish data alone with constant temperature \citep{tafalla2002}.
The new temperature profile is warmer in the outside and colder in the very centre and this results in a more peaked density structure
($n_0=2.1 \, 10^6$~\percc, $r_{0,n}=14^{\prime\prime}$, $\alpha_n=2.5$). The abundance 
rise in the centre of the core shown by  \citet{tafalla2002} is required also in our best fit although it is less steep ($\alpha_X=0.16$ vs 0.3).
\\
The reason for these changes in the temperature structure can be seen in the comparison between the observations and the predictions of the two models
presented in Fig~\ref{Fmc}. In each panel of Fig~\ref{Fmc} the data are shown in grey, the black--bold curves represent the prediction of our best fit model
while the thin--red lines show the best fit of the single dish data assuming constant temperature.
The lower temperature in the centre is required to fit the intensity of the \AMM (2,2) line which is highly over produced in the constant temperature
model (see panels g and l). Lowering the central temperature requires the external temperature to be increased in order to still 
fit the single dish observations of  \AMM (2,2) (panel c).
The quality of this new temperature profile can be checked also against the ratio of the \AMM(2,2) and the  \AMM(1,1)$_{hf}$ which is 
very sensitive to variations in the temperature profile (panel h). The constant temperature predictions are dramatically different from the observed ratio
of \AMM(2,2)/\AMM(1,1)$_{hf}$.
Adjustments to the abundance profile are finally needed to fit the intensity profile of the two observed \AMM(1,1) hyperfine components. The
characteristic profile of the  \AMM(1,1)$_{main}$ (panel e) and the flux loss of the central channels of the main component (panel k) are two 
sensitive indicators of a good fit to the abundance profile. In panel d the ratio between the two \AMM (1,1) hyperfine components shows the overall
quality of the adopted abundance profile. 
We note that the best fit of \citet{tafalla2002} better reproduces the intensity profile of the first red hyperfine component of  \AMM (1,1) 
observed with the single--dish (panel b and i). However increasing the intensity of the  \AMM(1,1)$_{hf}$  would require higher abundances or 
warmer inner temperatures and this would significantly worsen the modelling of the interferometric data (see panels d, f and h). 
A possible solution of the problem could be found introducing a 2D geometry \citep[see e.g. the rotationally flattened envelopes of][]{ulrich1976} 
but, since the effect on the temperature profile would be modest, this is beyond the scope of the present paper.

\section{Conclusions}\label{concl}
 We presented high-resolution, interferometric observations of \AMM \ and \AMMD \ toward L~1544, a 
low-mass starless core on the verge of forming a star. The gas temperature 
and deuterium fractionation are derived for the first time with a resolution
smaller than 800 AU. The main results of this work are:
\\
1) The gas temperature drops to $\sim$5.5 K at the core density
peak, in agreement with theoretical model predictions 
\citep{zucconi2001,evans2001,galli2002} and with the continuum modelling of \citet{doty2005}. 
In particular, our data can be  reproduced by the models of \citet{galli2002}, simultaneously 
fitting the dust continuum data. A similar conclusion about the gas temperature drop has been 
obtained by \citet{pagani2007} based on \NTHP, \NTDP,  and \AMM \ observations of \object{L~183}.
\\
2) The central density derived from literature data but including the temperature profile
reaches $2 \cdot 10^6$~\percc, 50\% higher than considering constant temperature at 8.75~K. 
\\
3) The column densities for both  \AMM \ and \AMMD \ 
are (factor of $\sim$2) larger than previously found with single-dish antennas
\citep{tafalla2002,shah2001}, evidence that we are probing a higher density 
gas and that, unlike CO,  the ammonia freeze--out is not significant
in the core centre, where $n(H_2) \ga 10^6$ cm$^{-3}$. 
\\
4) The deuterium fractionation in \AMM \ is 0.5$\pm$0.2,
almost constant in the central 4000 AU, where \AMMD\ is detected 
\\
5) The \AMMD\ map, unlike those of \AMM , peaks at the dust continuum peak showing that 
it is a very good tracer of the high density gas. A comparison between the velocity field extracted from \AMMD\ with the one
derived from \AMM \ shows an order of magnitude decrease of specific angular momentum from the large scales to the small scales.

\begin{acknowledgements}
We gratefully thank the IRAM-Plateau de Bure staff and NRAO-VLA  staff, and in particular Clemens Thum, 
Roberto Neri and Yancy Shirley, for support in preparing and performing these observations
A special thanks goes to Dr. Riccardo Cesaroni and Dr. Michiel Hogerijde for extensive help in the data reduction 
and analysis and to Jose Gon{\c c}alves for running a modified version of his temperature model.
A.C. acknowledges dr. Marcello Giroletti and dr. Lara Baldacci for interesting discussions that improved this 
manuscript. A.C. was supported by a fellowship from the European Research Training Network
"The Origin of Planetary Systems'' (PLANETS, contract number HPRN-CT-2002-00308)
at Leiden Observatory.

\end{acknowledgements}

\twocolumn

\clearpage

%
%


\end{document}